# Analysis of the Phase Separation Induced by a Free-Radical Polymerization in Solutions of Polyisobutylene in Isobornyl Methacrylate

**Ezequiel R. Soulé, Guillermo E. Eliçabe, Julio Borrajo and Roberto J. J. Williams[*]**

*Institute of Materials Science and Technology (INTEMA), University of Mar del Plata and National Research Council (CONICET), J. B. Justo 4302, 7600 Mar del Plata, Argentina*

* Corresponding author. E-mail: williams@fi.mdp.edu.ar. Tel: +54 223 481 6600. Fax: +54 223 481 0046.

**Abstract**

We analyze a polymerization-induced phase separation taking place in solutions of a polymer dissolved in a vinyl monomer undergoing an isothermal free-radical polymerization. The selected system was a solution of polyisobutylene (PIB) in isobornyl methacrylate (IBoMA). Cloud-point curves for PIBs of different molar masses were fitted with the Flory-Huggins equation stated for a 3-component system. Phase separation was analyzed for two different PIBs at off-critical compositions. Comparing experimental values of the volume fraction of dispersed phase with those predicted by the coexistence curves it was found that the system evolved close to equilibrium during an initial stage. But phase separation stopped at intermediate monomer conversions as assessed by scanning electron micrographs of partially converted materials. The evolution of the size distribution of dispersed domains was obtained from the fitting of light scattering (LS) spectra. Average sizes predicted by LS and measured in scanning electron micrographs were in rough agreement.

## 1. Introduction

When a polymerization is carried out in the presence of a second component (an oligomer, a polymer or a small molecule), a phase separation process usually takes place leading to different types of morphologies that depend on the initial composition and reaction conditions. Polymerization-induced phase separation (PIPS) is used in practice to synthesize a set of useful materials such as high-impact polystyrene (HIPS),[1] rubber-modified thermosets,[2] thermoplastic-thermoset blends,[3] polymer-dispersed liquid crystals,[4] thermally-reversible light scattering films,[5,6] nanostructured thermosets,[7] etc.

A thermodynamic analysis of PIPS shows that the main driving force for phase separation is the decrease in the absolute value of the entropic contribution to the free energy of mixing.[8] Other factors can also contribute to the demixing process such as a change in the interaction parameter due to changes in the chemical structure produced by polymerization, or the contribution of elastic energy after gelation (in the case of a thermosetting polymer).[9] For stepwise polymerizations the polymer may be considered as a single pseudo-component in a solution with the modifier. Cloud-point and coexistence curves can be predicted for this pseudo-binary system taking the polydispersities of the polymerizing component and the modifier into account.[9-11] However, for chainwise polymerizations the situation is more complex because at least three components are necessary for the thermodynamic analysis: the monomer, the polymer and the modifier. The simplest case is illustrated by the polymerization of a vinyl monomer (styrene, (meth)acrylic monomer, etc.) in the presence of a modifier, leading to a linear polymer. Several recent papers have analyzed PIPS in this type of systems, e.g., a polyethylene wax dissolved in styrene[12] or in isobornyl methacrylate,[13] polystyrene in 2-chlorostyrene,[14,15] poly(dimethylsiloxane-*co*-diphenylsiloxane) in 4-chlorostyrene[16,17] or in *n*-butyl methacrylate.[18] In all these references phase separation at a specific temperature is qualitatively discussed in a triangular diagram using a single coexistence curve assumed equal to the cloud-point curve. But polydispersities of the modifier and the generated polymer produce a shift of coexistence curves with monomer conversion, a fact that has not been previously analyzed in the literature for chainwise polymerizations.

The aim of this study was to illustrate the shift of coexistence curves with monomer conversion in a solution of polyisobutylene (PIB) in isobornyl methacrylate (IBoMA) that produces poly(isobornyl methacrylate) (PIBoMA) through a free-radical polymerization. We will show that the description of the phase separation process can be significantly modified when the analysis is based in coexistence curves instead of the schematic use of the cloud-point curve. Two particular formulations were selected to illustrate the usefulness of combining thermodynamic predictions with experimental results obtained by scanning electron microscopy (SEM) and light scattering (LS), to analyze a phase separation induced by the free-radical polymerization.

## 2. Experimental Section

**2.1 Materials.** Isobornyl methacrylate (IBoMA, Aldrich) was used as received. It contained 150 ppm of *p*-methoxyphenol (MEHQ, methyl ether hydroquinone) as inhibitor. Its mass density at 15 °C was $\rho_{IBoMA} = 0.983$ g/cm$^3$ and its refractive index was $n_{IBoMA} = 1.48$. Benzoyl peroxide (BPO, Akzo-Nobel) was used as initiator. Poly(isobornyl methacrylate) (PIBoMA) was synthesized from IBoMA using 2 wt % BPO. Approximately 2 g of the IBoMA-BPO solution were heated in a closed glass tube at 80 °C for 1 h, followed by 30 min at 140 °C to complete the reaction. The molar mass distribution of the synthesized PIBoMA was determined by size exclusion chromatography using a universal calibration curve as reported elsewhere.[19] Average values were $M_n = $ 1.71 x 10$^5$, $M_w = $ 1.01 x 10$^6$, $M_z = $ 3.41 x 10$^6$. Mass density of PIBoMA at 15 °C was $\rho_{PIBoMA} = $ 1.06 g/cm$^3$.

Three commercial poly(isobutylenes) (PIB, Repsol YPF, Argentina), were used. Molar mass distributions were reported by the supplier. Molar mass averages and other characteristics are reported in Table 1.

**2.2 Cloud-point Curves.** Experimental cloud-point curves for ternary systems of PIB (025, 5 or 30), IBoMA and PIBoMA, at 80 °C, were reported in a previous publication,[19] and will be used here for modeling purposes. Cloud-point curves were obtained for physical mixtures of the three components and in the course of polymerization starting from solutions of PIB in IBoMA. Experimental details of these determinations were reported in the previous publication.[19]

Cloud-point curves for selected binary and ternary solutions were determined in the present study to obtain values of the interaction parameters. Physical blends of two or the three components were placed between two glass slides separated by a 0.5 mm stainless steel spacer, and the cloud-point temperature was determined using a Leica DMLB microscope provided with a videocamera (Leica DC100), a temperature-controlled stage (Linkam THMS 600). Blends were heated to a temperature above the cloud-point curve and then cooled at rates comprised between 0.2 and 2 °C/min until the cloud-point was observed. The procedure was performed several times observing good repeatability of the cloud-point temperature at different cooling rates.

**2.3. Scanning Electron Microscopy (SEM).** In order to obtain SEM micrographs of partially converted samples it was necessary that their glass transitions temperatures had increased to values higher than room temperature. This fixed a minimum monomer conversion level for SEM observations of partially reacted samples. Two formulations were selected: a) 15 wt % PIB30 in IBoMA and b) 30 wt % PIB5 in IBoMA. The former exhibited a cloud-point conversion, $p_{cp}$ = 0.41 and could be manipulated to obtain SEM micrographs for monomer conversions $p > 0.55$; the latter had a cloud-point conversion, $p_{cp}$ = 0.425 and it was required to attain monomer conversions $p > 0.60$ for SEM observations. Therefore, there was an initial stage of the phase separation process where no information could be obtained by SEM. Polymerizations were carried out in glass tubes of 3-mm diameter to assure an isothermal reaction. A set of tubes filled with the appropriate blend was placed in a thermostat held at 80 ºC. In order to transform polymerization times into monomer conversions we used conversion vs. time curves obtained from kinetic results reported in our previous publication.[19] The main problem was the appearance of an induction time related to the consumption of the inhibitors initially present in the formulation. And the duration of this induction period depends on variables that are very difficult to reproduce such as the temperature history and the concentration of dissolved oxygen. To circumvent this problem we fixed a point in the monomer conversion scale by assigning the cloud-point conversion previously determined to the cloud-point time observed in the set of glass tubes. Then the time scale was transformed into a monomer conversion scale using the polymerization kinetics.[19] This enabled us to select appropriate monomer conversion values to extract tubes from the bath and stop the reaction by a quenching to 0 ºC. Samples were extracted from the tubes and fractured. SEM micrographs of fracture surfaces coated with a fine gold layer were obtained using a Jeol JSM 6460 LV device.

**2.4. Light Scattering (LS).** The light scattering apparatus was built up using a Fraunhofer configuration consisting of a linear array of photodiodes placed to detect the light scattered by a thin sample illuminated by a 17 mW He-Ne laser with random polarization. The laser beam was attenuated by a neutral density filter, expanded by a 5x beam expander and trimmed by an iris diaphragm before reaching the sample. The sample holder was made of two glass windows separated by a 1.2 mm spacer. It was placed in an aluminum block provided with a computer-controlled electrical heating system. The light scattered from the sample was collected by 14 cathode monolithic silicon photodiode linear arrays of 16 elements each, manufactured by Photonics Detectors Inc. The elements, of length 25.20 mm each, were located one next to the other with a small gap of 0.25mm. The active area of each element was $2.31\text{mm}^2$. Every photodiode operated as a current source linearly controlled by the light intensity. An opal diffusing glass was used to calibrate the photodiodes individually and thus compensate for their different gains. The current from the photodiode circulated through a resistance generating a voltage proportional to the incident light. This voltage was sensed by 29 integrated circuits, consisting of an 8-channel multiplexer and a 12-Bit A/D converter. Serial communication was used to receive the control signals employed to select the channel that was sampled. Signals were sent to a microprocessor that carried out all the coordination and control tasks, connected to a personal computer through a serial port. This type of connection imposed a minimum sampling time of 1 s, although the time needed to collect a whole spectrum was much less.

The same two formulations selected to obtain SEM micrographs of partially reacted samples were used to obtain LS spectra in the course of phase separation. The cloud-point time was evidenced by the beginning of an increase in the intensity of scattered light. The knowledge of the cloud-point conversion that was determined independently, enabled to transform the time scale into a monomer conversion scale using the polymerization kinetics.[19]

## 3. Results and Discussion

**3.1 Thermodynamic Analysis of Cloud-Point Curves.** Figure 1 shows experimental cloud points for the three ternary systems at 80 ºC. Full symbols are experimental points obtained for physical blends while unfilled symbols represent values obtained during polymerization. These experimental points were reported in the previous study[19] (two values were re-measured and slightly shifted).

In the rectangular triangle, IBoMA is placed in the upper vertex, PIBoMA in the lower vertex at the left and PIB in the lower vertex at the right. The region bounded by the dashed line located close to the vertex of pure PIBoMA includes compositions that are in the glassy state at 80 ºC. The initial solution of IBoMA and PIB is located in the diagonal of the triangle and the polymerization is represented by a vertical line starting in the diagonal and ending in the base of the triangle (or somewhere in the glassy region if the reaction is arrested by vitrification). When the trajectory reaches the cloud-point curve corresponding to the specific PIB, phase separation begins to take place. Experimental results show that PIB30 is less miscible than PIB5 which in turn is much less miscible than PIB025, in agreement with what could be expected from the molar mass distributions reported in Table 1 (miscibility decreases with an increase in the molar mass).

In every case the cloud point was observed at the same composition (monomer conversion) for physical blends and during the polymerization-induced phase separation. In the case of physical blends a very slow cooling rate was used enabling the system to produce phase separation when reaching the metastable region of the phase diagram. The fact that the same cloud-point was obtained during polymerization is explained by the slow polymerization rate (compared to the phase separation rate) at the selected reaction conditions. Therefore, we can assess that PIPS did also take place as soon as the system entered the metastable region. In turn, this means that for off-critical compositions phase separation should occur by a nucleation-growth-coarsening mechanism.

Another factor that could have modified the location of cloud-point curves measured for physical blends and in the course of polymerization is the grafting of PIB by PIBoMA chains. This would have increased the miscibility window in blends produced by chemical reaction. However, experimental determinations showed that for our system grafting was not significant.[19] It should be also pointed out that the molar mass distribution of PIBoMA could be affected by varying the amount and type of PIB in the initial solution. However, for the selected experimental conditions unimodal molar mass distributions of PIBoMA located in the range of high molar masses, were observed in every case.[19] In the range of high molar masses small changes in the molar mass distributions have no significant effect on miscibility (negligible contribution of the corresponding entropic terms to the free energy of mixing).

Experimental cloud-point conversions were fitted with the Flory-Huggins (FH) model with an excess free energy accounted by interaction parameters that can depend both on composition and temperature. The free energy per unit volume, $\Delta G$, of a blend of IBoMA (subscript 0), PIB (subscript 1), and PIBoMA (subscript 2), is expressed by the following equation:

$$(V_r/R_g T)\Delta G = (\phi_0/r_0)\ln\phi_0 + \Sigma(\phi_{1i}/r_{1i})\ln\phi_{1i} + \Sigma(\phi_{2i}/r_{2i})\ln\phi_{2i} +$$

$$g_{01}\phi_0\phi_1 + g_{02}\phi_0\phi_2 + g_{12}\phi_1\phi_2 \qquad (1)$$

$V_r$ is an arbitrary reference volume that in this case was taken equal to 1 cm$^3$/mol, $R_g$ is the gas constant, $T$ is the absolute temperature, $\phi$ indicates the volume fraction of IBoMA (subscript 0), of a particular *i*-mer of PIB (subscript 1i), and of a particular *i*-mer of PIBoMA (subscript 2i). Volume fractions were calculated from the mass fraction of IBoMA, the molar mass distributions of PIB and PIBoMA expressed into mass fractions, and using the corresponding mass densities. Total volume factions of components 1 and 2 are defined as $\phi_1 = \Sigma\phi_{1i}$ and $\phi_2 = \Sigma\phi_{2i}$. The size of every species is defined by the factor *r* defined as the ratio of the volume fraction of a particular species with respect to the reference volume. The experimental continuous molar mass distributions of PIB and PIBoMA were transformed into discrete distributions containing 20-30 species, checking that resulting values of $M_n$, $M_w$ and $M_z$ fitted the experimental values within 0.1 %.

The first three terms of the right-hand side of the FH equation represent the combinatorial contribution to free energy (entropic contribution) while the last three terms represent the excess contribution to free energy expressed in terms of the binary interaction parameters, $g_{ij}$. When necessary these binary interaction parameters were taken as a function of both composition and temperature.

Chemical potentials of the three components may be obtained from eq 1 by standard procedures,[20] written in terms of separation factors:

$$\sigma_0 = (1/r_0)\ln(\phi_0^\beta/\phi_0^\alpha)$$
$$\sigma_1 = (1/r_{1i})\ln(\phi_{1i}^\beta/\phi_{1i}^\alpha) \qquad (2)$$
$$\sigma_2 = (1/r_{2i})\ln(\phi_{2i}^\beta/\phi_{2i}^\alpha)$$

In eq 2, the superscript $\beta$ represents the emergent phase while the superscript $\alpha$ indicates the original phase. Equating the chemical potentials of the three components in both phases leads to a set of three algebraic equations. A fourth equation states that the sum of volume fractions of species in the emergent phase must be equal to 1. A set of four equations in four unknowns (three separation factors and the cloud-point temperature) is generated. A unique solution may be obtained provided the binary interaction parameters are known. Therefore the following step was the determination of these parameters.

Figure 2a-c shows experimental cloud-point temperatures ($T_{cp}$) obtained for the binaries IBoMA-PIB, exhibiting an upper-critical-solution-temperature (UCST) behavior. A fitting with the FH equation (eq 1 reduced to the 0-1 binary) was performed using the following functionality for the interaction parameter, proposed by Prausnitz and coworkers:[21]

$$g_{01} = (a_{01} + b_{01}/T)/[1/c_{01}(1 - \phi_1)]\ln[(1 - c_{01}\phi_1)/(1 - c_{01})] \qquad (3)$$

For every PIB the best values of the three coefficients defined in eq 3 were searched by minimizing $\Sigma[T_{cp} \text{ (predicted)} - T_{cp} \text{ (exp)}]^2$. The Levenberg-Marquardt algorithm included in Mathcad 2001 was used for this purpose. The resulting set of coefficients is shown in Table 2 for the three PIB. The curves plotted in Figure 2a-c represent the fitting of experimental points with the FH model.

Regarding the two other couples, IBoMA and PIBoMA were completely soluble while PIB5 and PIB30 did not exhibit complete miscibility with PIBoMA at any temperature in the range that could be explored to avoid degradation. A cloud-point curve could however be obtained for the couple PIB025-PIBoMA, as shown in Figure 3. Cloud-point curves were also obtained adding 5 and 8 wt % IBoMA to the PIB025-PIBoMA blends, as is also shown in Figure 3. A small addition of the monomer produced a large increase in the miscibility of both polymers.

The binary PIB025-PIBoMA was modeled using an interaction parameter defined by:

$$g_{12} = (a_{12} + b_{12}/T)/[1/c_{12}(1 - \phi_2)]\ln[(1 - c_{12}\phi_2)/(1 - c_{12})] \quad (4)$$

For ternary blends of IBoMA-PIB025-PIBoMA, the interaction parameter between IBoMA and PIBoMA was defined by:

$$g_{02} = a_{02} + b_{02}/T \quad (5)$$

Best fitting parameters were searched simultaneously for the binary and ternary blends using a similar procedure as for the IBoMA-PIB couples. The $g_{01}$ function with the parameters shown in Table 2 was used for the fitting of the ternary blend. The arbitrary extension of the use of this function to a ternary blend (the effect of composition is assumed the same in a binary and in a ternary blend) is self-compensated when searching the best functions for $g_{12}$ and $g_{02}$. In this way, the FH equation for the ternary blend reduces correctly to the binary systems.

Values obtained for the PIB025-PIBoMA couple are shown in Table 2. Best values for the IBoMA-PIBoMA couple were $a_{02} = 0.00609$, $b_{02} = -1.602$. For this binary system a lower-critical solution temperature (LCST) behavior is predicted with a critical temperature close to 140 °C. The fitting of experimental points is shown by the full curves of Figure 3.

Now, eq 1 may be used to calculate cloud-point curves of the ternary blends at 353 K, using the interaction parameters already found. For the couples PIB5-PIBoMA and PIB30-PIBoMA the same value of $c_{12}$ as that found for PIB025-PIBoMA, was used (Table 2). The factor $(a_{12}+b_{12}/353)$ was left as the only fitting parameter of the ternary blends containing PIB5 or PIB30. Results for the best fitting are indicated in Table 2. The curves plotted in Figure 1 represent the final fitting of cloud-point curves achieved by the FH model.

**3.2 Coexistence Curves During Polymerization.** The evolution of coexistence curves during polymerization may be predicted from eq 1 by selecting the starting formulation. The polymerization-induced phase separation for two different formulations will be analyzed in what follows: a) 15 wt % PIB30 in IBoMA and b) 30 wt % PIB5 in IBoMA. Cloud-point conversions ($p_{cp}$) may be calculated from the curves plotted in Figure 1. They are: $p_{cp} = 0.41$ for the blend with 15 wt % PIB30 and $p_{cp} = 0.425$ for the blend with 30 wt % PIB5. For $p > p_{cp}$ the evolution of the system will be bounded by the coexistence curves. If equilibrium is attained at every monomer conversion, the composition and volume fraction of both phases can be obtained from the coexistence curves.

Coexistence curves were determined by equating chemical potentials of the three components in both phases, similarly to the calculation of cloud-point curves, and incorporating the volume fraction of the emergent phase as a new parameter. The procedure used to solve the corresponding equations may be found in the literature.[22] The critical point was also calculated using equations developed for ternary blends containing polydisperse components adapted to consider interaction parameters depending on composition and temperature.[22]

Figures 4a and 4b show the evolution of the equilibrium compositions of both phases for the selected blends. Critical points are located at much higher PIB concentrations than those used in the selected blends. Therefore, phase separation takes place at off-critical conditions.

Coexistence curves predict that the emergent phase ($\beta$) is a binary solution of PIB in IBoMA without any significant amount of PIBoMA. However, a large fraction of PIB remains in solution in the continuous phase, a result that could not have been predicted by analyzing the phase separation process only with the cloud-point curve. This fractionation effect is typical of oligomers exhibiting a broad molar mass distribution. A high molar mass fraction is segregated to the emergent phase while a low molar-mass fraction remains miscible in the IBoMA-PIBoMA solution. As PIBoMA species have high molar masses they are not fractionated and remain in the continuous phase. Figure 5 shows the variation of the weight average molar mass of the PIB fraction in the $\alpha$ (continuous) and $\beta$ (emergent) phases after the cloud point, for the blend containing 15 wt % PIB30. The $M_\mathrm{w}$ value is higher in the emergent phase than in the continuous phase. As a result of this fractionation the solubility of PIB in the continuous phase varies slowly and a high residual fraction is predicted at the conversion where vitrification takes place.

Figure 4 also shows that the monomer is fractionated between both phases keeping a higher mass fraction in the continuous phase than in the emergent phase. Assuming that a fraction of the initiator is also fractionated to the emergent phase the monomer can continue polymerization producing PIBoMA inside the emergent phase (with a different rate than in the matrix due to changes in composition). This will produce a secondary phase separation process inside the primary emergent phase leading to an IBoMA-PIBoMA emergent phase. This new phase can be located inside the primary dispersed domains (salami or core-shell structure) or can diffuse away and join the continuous phase. In our previous kinetic study we found evidence of polymerization proceeding in two different phases only when using a high mass fraction of PIB (50 wt %) in the initial formulation.[19] In the present study we did not find any experimental evidence of the polymerization inside the emergent phase.

At high conversions the primary phase separation rate will slow down due to the high viscosity of the system and it will finally be arrested by vitrification. In this case the composition of the continuous phase will depart from the coexistence curve approaching a vertical line. The comparison between experimental and predicted evolutions of the volume fraction of dispersed phase can be used to discuss this possibility.

**3.3 Evolution of the Volume Fraction of Dispersed Phase.** SEM micrographs were obtained for the blend with 15 wt % PIB30 in the monomer conversion range comprised between $p = 0.56$ and $p = 0.84$ (Figure 6), and for the blend with 30 wt % PIB5 in the monomer conversion range lying between $p = 0.63$ and $p = 0.88$ (Figure 7). Unfortunately it was not possible to explore a range closer to the cloud-point conversions due to the problems discussed in the Experimental Section.

For both blends SEM micrographs show the presence of a relatively diluted dispersion of small spherical particles. These particles correspond to the emergent ($\beta$) phase, rich in PIB, produced by a nucleation-growth-coarsening mechanism. The volume fraction of dispersed phase ($V_D$) was estimated from the fraction of area occupied by particles ($A_D$), corrected by assuming that the crack propagation plane has a height equal to the number-average diameter of the particles, $D_{av}$. This leads to:[23]

$$V_D = (2/3) A_D (D_{av})^2/(D^2)_{av} \qquad (6)$$

Figures 8a and 8b show a comparison of the volume fraction of dispersed phase calculated from the thermodynamic model assuming equilibrium conditions, and the experimental values determined by SEM for the two selected blends. For both blends, the volume fraction lies close to the thermodynamic prediction up to monomer conversions close to 0.65. But at higher monomer conversions phase separation seems to be arrested as the volume fraction of dispersed phase does not exhibit any significant change. This is possibly due to the high viscosity of the continuous phase at these monomer conversion levels.

Therefore, the phase separation process seems to proceed at a fast initial rate by a nucleation-growth-coarsening process, driving the compositions to equilibrium. As monomer conversion increases so does the viscosity of the continuous phase. This slows the phase separation rate and produces a departure from the equilibrium curve. For monomer conversions higher than about 0.65 no significant changes are observed.

**3.4 Evolution of the Particle Size Distribution.** Light scattering (LS) spectra were obtained in situ for the two selected blends polymerized at 80 °C. Figures 9a and 9b show the spectra obtained for both blends at increasing monomer conversion values. The presence of a maximum in LS spectra for a polymerization-induced phase separation occurring by a nucleation-growth-coarsening process has been discussed in the literature.[24,25]

Even though the cloud point was detected as a first significant increase in the scattering intensity, spectra could only be modeled starting from higher monomer conversions due to the high noise/signal ratio of the spectra obtained close to the cloud point. Corresponding monomer conversions were 0.488 for the blend with 15 wt % PIB30 and 0.512 for the blend with 30 wt % PIB5. This enlarged significantly the monomer conversion range that could be monitored by SEM (starting at 0.56 for the blend with 15 wt % PIB30 and 0.63 for the blend with 30 wt % PIB5).

In order to get the evolution of particle size distributions from the spectra of Figure 9, a light scattering model proposed by Pedersen[26] was used. The model is based on the local monodisperse approximation that assumes that the positions of the particles are completely correlated with their sizes. This approximation realizes the system as many subsystems in which the particles are monodisperse. Pedersen showed that this approach worked correctly even for large polydispersities.[26] The intensity of scattered light ($I_d$) is calculated as the sum of the scattering from the subsystems weighted according to the particle size distribution:

$$I_d(q) \sim \int_0^\infty f(R) S(q,R) F^2(q,R) dR \qquad (7)$$

In eq 7, $q = (4\pi n_m/\lambda_0)\sin(\theta/2)$ is the modulus of the scattering vector where $n_m$ is the refractive index of the medium (taken as 1.485 for the blend with 15 wt % PIB30 and 1.489 for the blend with 30 wt% PIB5), $\lambda_0$ is the wavelength of incident light in vacuum (632.8 nm) and $\theta$ is the scattering angle. The form factor, $F(q,R)$, for a sphere of radius $R$ is given by:

$$F(q,R) = (\Delta n/q^3)(\sin qR - qR \cos qR) \qquad (8)$$

where $\Delta n$ is the absolute value of the difference between the refractive indices of the medium and the particles. The structure factor $S(q,R)$ was calculated by the expression given by Pedersen in the Percus-Yevick approximation.[26]

A log-normal function was selected to fit the particle size distribution:

$$f(R) = (s/\pi R^2)^{1/2} \exp[-s \log^2(R/R_0)] \qquad (9)$$

The LS model contains 4 adjustable parameters: the proportionality constant of eq 7 that depends on the particle concentration and is lumped with $(\Delta n)^2$ present in the form factor, an effective parameter present in the structure factor that describes the interference effects,[26] and the two parameters ($s$ and $R_0$) characterizing the particle size distribution. The best set of the 4 adjustable parameters was obtained for every one of the selected spectra. An excellent fitting of LS spectra was obtained as shown by the smooth curves in Figures 9a and 9b, except for a region close to $q = 2$ $\mu m^{-1}$.

Figures 10a and 10b show the evolution of particle size distributions calculated by the model for the two blends. For the blend with 15 wt % PIB30 there is a significant shift of the particle size distribution to the right, in the monomer conversion range comprised between 0.488 and 0.578. This indicates that growth-coarsening mechanisms are still active in this range. For monomer conversions higher than 0.61 there is practically no further evolution of the particle size distribution meaning that phase separation was practically arrested when the system reached this monomer conversion. This is in excellent agreement with the evolution of the volume fraction of dispersed phase shown in Figure 8a.

For the blend with 30 wt % PIB5 there is also a shift of particle size distributions to the right, in the range of monomer conversions comprised between 0.512 and 0.58. For monomer conversions higher than 0.62 there was no significant evolution of the particle size distribution in agreement with the evolution of the volume fraction of dispersed phase shown in Figure 8b.

Figures 11a and 11b show a comparison of average sizes calculated from the log-normal distribution with experimental values obtained from SEM micrographs. A reasonable agreement was obtained evidencing the predictive capability of Pedersen's model.

It is interesting to observe that there was still an evolution of the spectra of scattered light plotted in Figures 9a and 9b in the monomer conversion range where there was no evolution of the particle size distribution and the volume fraction of dispersed phase (6 spectra in Figure 9a and 7 spectra in Figure 9b correspond to this situation). The only adjustable parameter that varied in this range was the proportionality constant of eq 7 that includes $(\Delta n)^2$. The increase in the intensity of LS spectra in the range of high monomer conversions is assigned to an increase in the difference of refractive indices between the particles and the matrix due to the continuous purification of both phases in the corresponding neat components.

## 4. Conclusions

A thermodynamic analysis of phase separation during a free-radical polymerization must take into account the polydispersity of the modifier, particularly if it is a low molar mass polymer (PIB, polyethylene waxes,[12,13] random copolymers of dimethyl and diphenylsiloxane[16-18]). In this situation, coexistence curves differ significantly from the initial cloud-point curve due to the fractionation of the modifier between both phases. Predicted coexistence curves can be used to analyze the phase separation process along polymerization. In the present study two particular formulations were selected in the off-critical range leading to a dispersion of particles rich in PIB. It was found that phase separation took place at a fast rate (relative to the polymerization rate) when the system entered the metastable region at conversions close to 0.40. The evolution of the volume fraction of dispersed phase and the particle size distribution was analyzed using the predicted coexistence curves together with SEM and LS results. It was found that a primary phase separation took place in the 0.40 – 0.60 monomer conversion range by a nucleation-growth-coarsening mechanism, carrying the composition of phases close to the values predicted by the coexistence curves (as inferred from the agreement of experimental and predicted values of the volume fraction of dispersed phase at the end of this interval). However, the primary phase separation became slower and was finally arrested at monomer conversions close to 0.65, a fact explained by the increase in the viscosity of the matrix. But the evolution of LS spectra showed that the difference in the refractive indices of both phases increased continuously for monomer conversions higher than 0.65 due to the continuous purification of both phases in the corresponding neat components.

**Acknowledgment**

We acknowledge the financial support of the following institutions of Argentina: University of Mar del Plata, National Research Council (CONICET), and National Agency for the Promotion of Science and Technology (ANPCyT).

**Table 1. Denomination, Molar Mass Averages, Mass Densities at 15 ºC, and Refractive Indices of the Commercial Polyisobutylenes**

| Polyisobutylene | $M_n$ | $M_w$ | $M_z$ | $\rho_{PIB}$ (g/cm$^3$) | $n_{PIB}$ |
|---|---|---|---|---|---|
| PIB025 | 674 | 1033 | 1490 | 0.875 | 1.51 |
| PIB5 | 808 | 1912 | 5164 | 0.888 | 1.51 |
| PIB30 | 1339 | 3035 | 6879 | 0.904 | 1.51 |

**Table 2. Parameters of the Binary Interaction Coefficients $g_{01}$ and $g_{12}$**

| Polyisobutylene | $a_{01}$ | $b_{01}$(K) | $c_{01}$ | $a_{12}$ | $b_{12}$(K) | $a_{12}+b_{12}/353$ | $c_{12}$ |
|---|---|---|---|---|---|---|---|
| PIB025 | -0.00269 | 1.733 | 0.279 | -0.000536 | 0.3344 | 0.000411 | 0.532 |
| PIB5 | 0.00163 | 0.2373 | 0.668 | | | 0.000530 | 0.532 |
| PIB30 | 0.00186 | 0.1700 | 0.651 | | | 0.000595 | 0.532 |

**Legends to the Figures**

**Figure 1.** Cloud-point curves for the three ternary systems at 80 °C. Full symbols are experimental points obtained for physical blends while unfilled symbols represent values obtained during polymerization. Curves represent the fitting obtained with the thermodynamic model.

**Figure 2.** Cloud-point curves for the binary IBoMA-PIB systems; (a) PIB025, (b) PIB5, (c) PIB30. Symbols are experimental points and the curves represent the fitting obtained with the functions proposed for the binary interaction parameters.

**Figure 3.** Cloud-point curves for the binary PIB025-PIBoMA and for ternary blends containing 5 and 8 wt % IBoMA. Symbols are experimental points and the curves represent the fitting obtained with the functions proposed for the binary interaction parameters.

**Figure 4.** Composition of phases at equilibrium during polymerization; (a) 15 wt % PIB30, (b) 30 wt % PIB5.

**Figure 5**. Variation of the weight average molar mass of the PIB fraction in the $\alpha$ (continuous) and $\beta$ (emergent) phases after the cloud point for the blend containing 15 wt % PIB 30.

**Figure 6.** SEM micrographs for the blend with 15 wt % PIB30 obtained at different monomer conversions; (a) $p = 0.56$, (b) $p = 0.84$.

**Figure 7.** SEM micrographs for the blend with 30 wt % PIB5 obtained at different monomer conversions; (a) $p = 0.63$, (b) $p = 0.88$.

**Figure 8.** Volume fraction of dispersed phase predicted by the thermodynamic model assuming instantaneous equilibrium (curve) compared with experimental values obtained by SEM; (a) 15 wt % PIB30, (b) 30 wt% PIB5.

**Figure 9.** Light scattering spectra plotted as intensity (arbitrary units) as a function of the modulus of the scattering vector $q$. Non-smooth curves are experimental results and smooth curves represent the fitting of the LS model; (a) 15 wt % PIB30 for the following monomer conversions: 0.488, 0.541, 0.578, 0.61, 0.64, 0.667, 0.713, 0.753, 0.827; (b) 30 wt % PIB5 for the following monomer conversions: 0.512, 0.536, 0.558, 0.58, 0.601, 0.621, 0.641, 0.66, 0.678, 0.712, 0.788, 0.881.

**Figure 10.** Evolution of particle size distributions calculated by the LS model for the two blends; (a) 15 wt % PIB30 for the following monomer conversions: 0.488, 0.541, 0.578, 0.61, 0.813; (b) 30 wt % PIB5 for the following monomer conversions: 0.512, 0.536, 0.558, 0.621 and 0.80.

**Figure 11.** Comparison of average sizes calculated from the log-normal distribution with experimental values obtained from SEM micrographs; (a) 15 wt % PIB30, (b) 30 wt % PIB5.

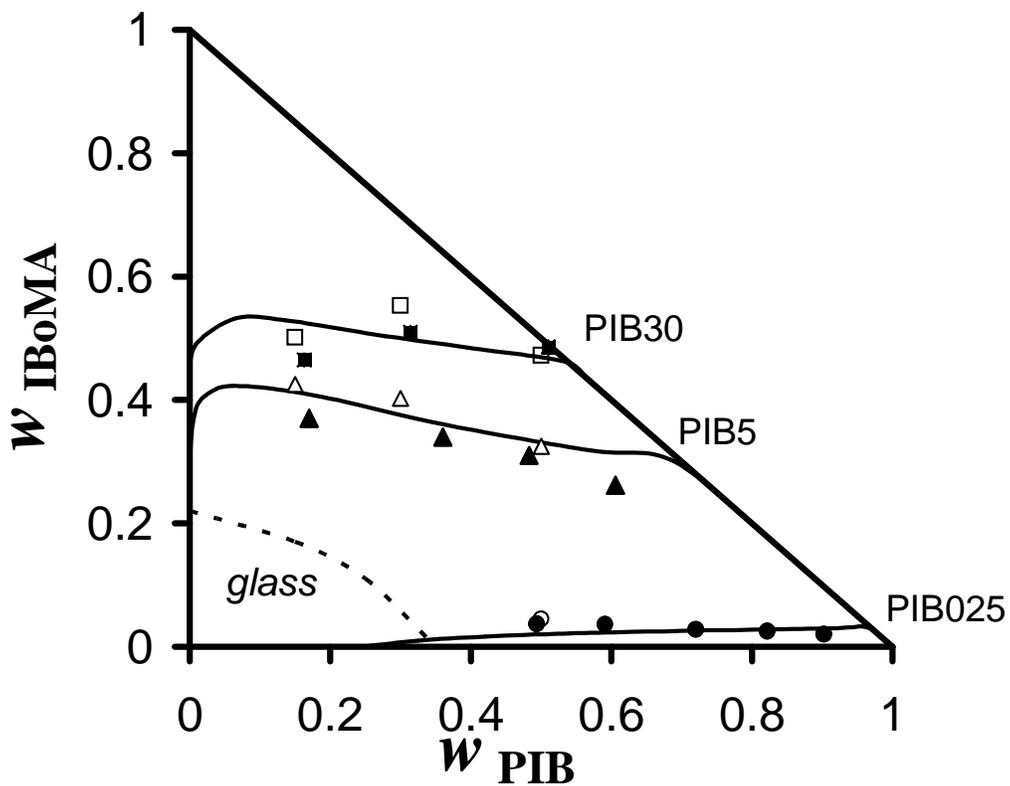

**Figure 1.** Cloud-point curves for the three ternary systems at 80 °C. Full symbols are experimental points obtained for physical blends while unfilled symbols represent values obtained during polymerization. Curves represent the fitting obtained with the thermodynamic model.

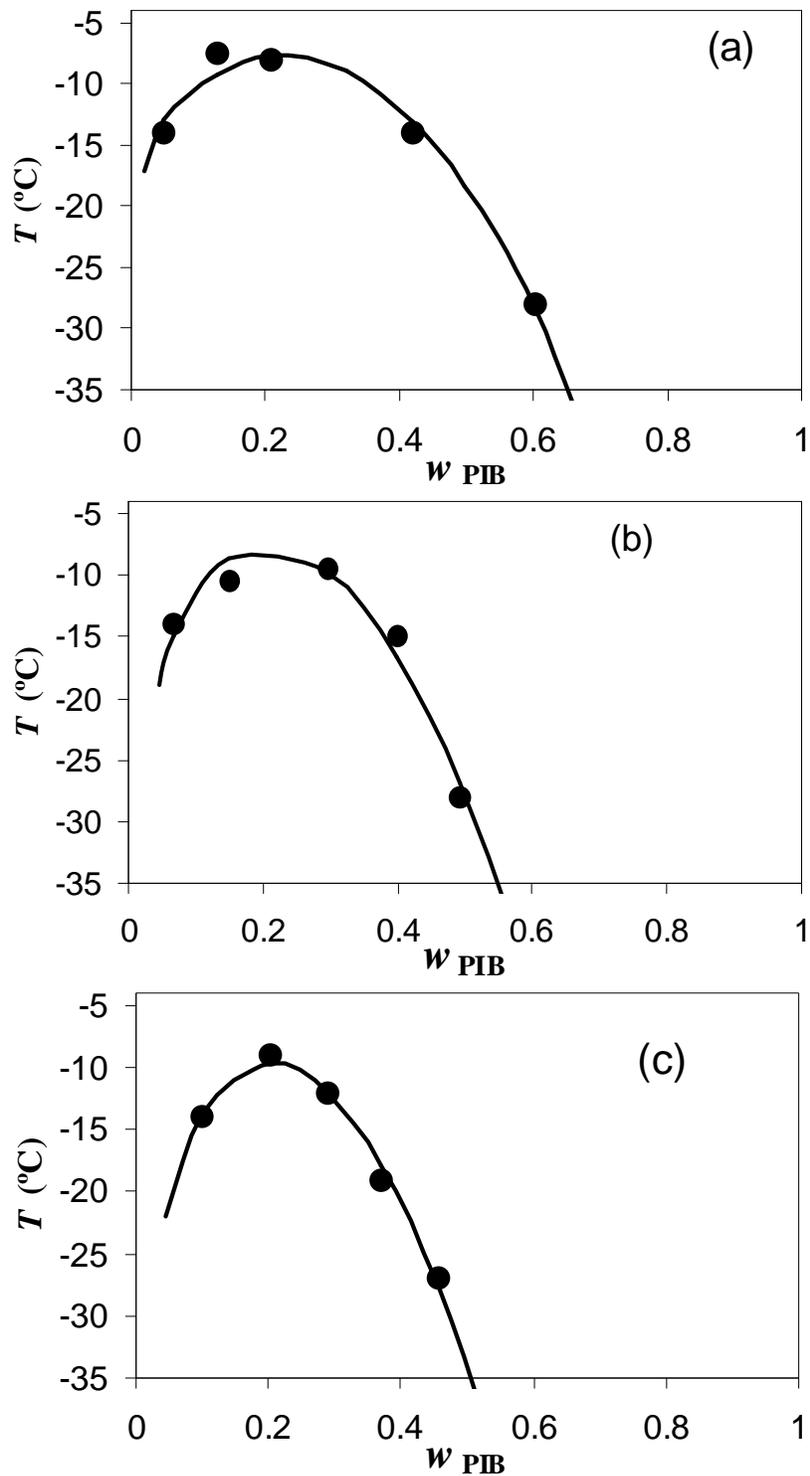

**Figure 2.** Cloud-point curves for the binary IBoMA-PIB systems; (a) PIB025, (b) PIB5, (c) PIB30. Symbols are experimental points and the curves represent the fitting obtained with the functions proposed for the binary interaction parameters.

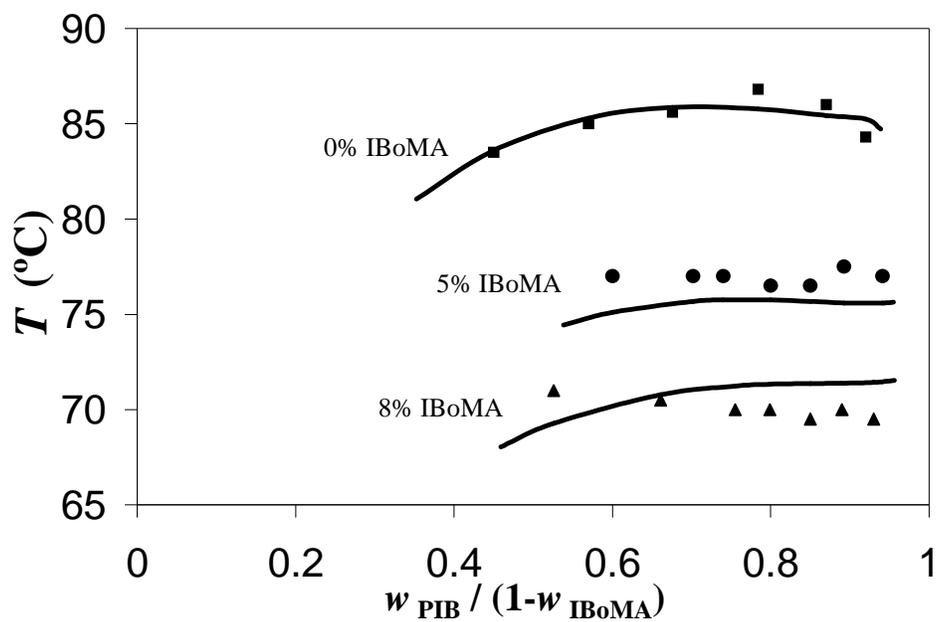

**Figure 3.** Cloud-point curves for the binary PIB025-PIBoMA and for ternary blends containing 5 and 8 wt % IBoMA. Symbols are experimental points and the curves represent the fitting obtained with the functions proposed for the binary interaction parameters.

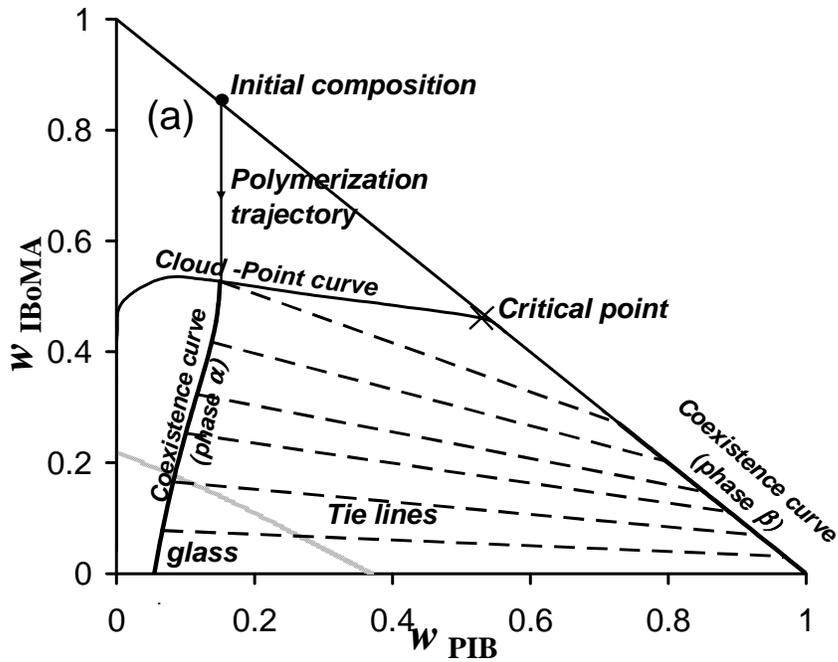

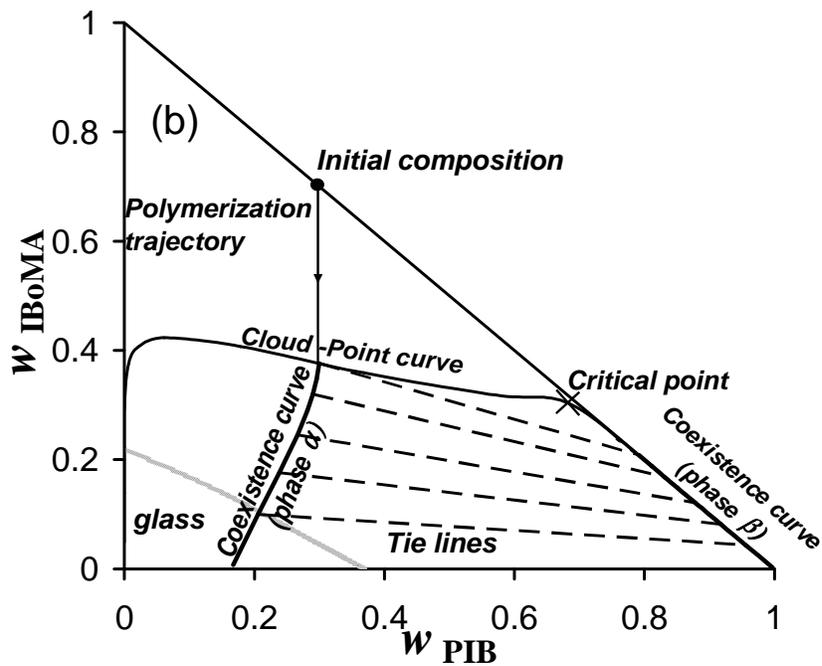

**Figure 4.** Composition of phases at equilibrium during polymerization; (a) 15 wt % PIB30, (b) 30 wt % PIB5.

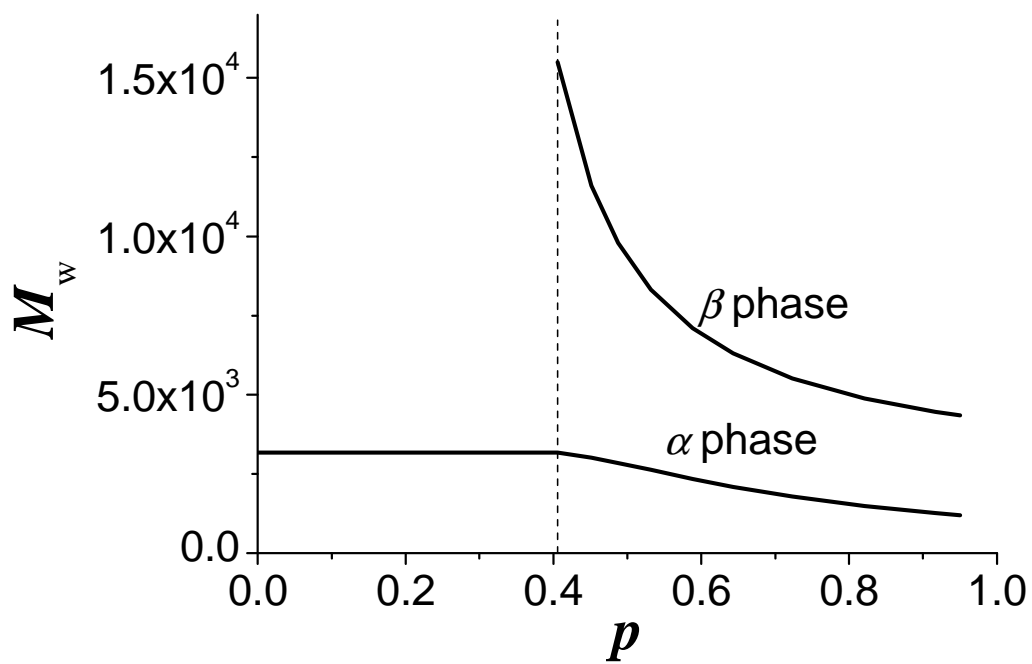

**Figure 5**. Variation of the weight average molar mass of the PIB fraction in the $\alpha$ (continuous) and $\beta$ (emergent) phases after the cloud point for the blend containing 15 wt % PIB 30.

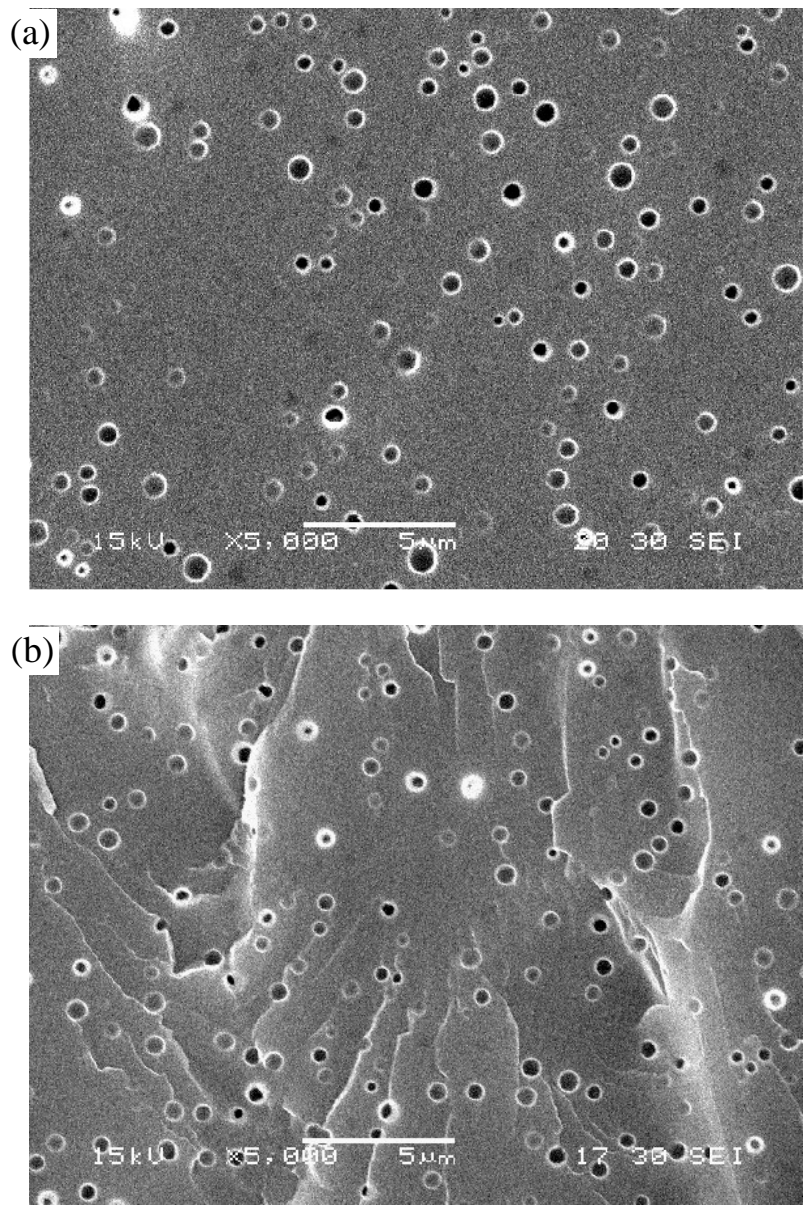

**Figure 6.** SEM micrographs for the blend with 15 wt % PIB30 obtained at different monomer conversions; (a) $p = 0.56$, (b) $p = 0.84$.

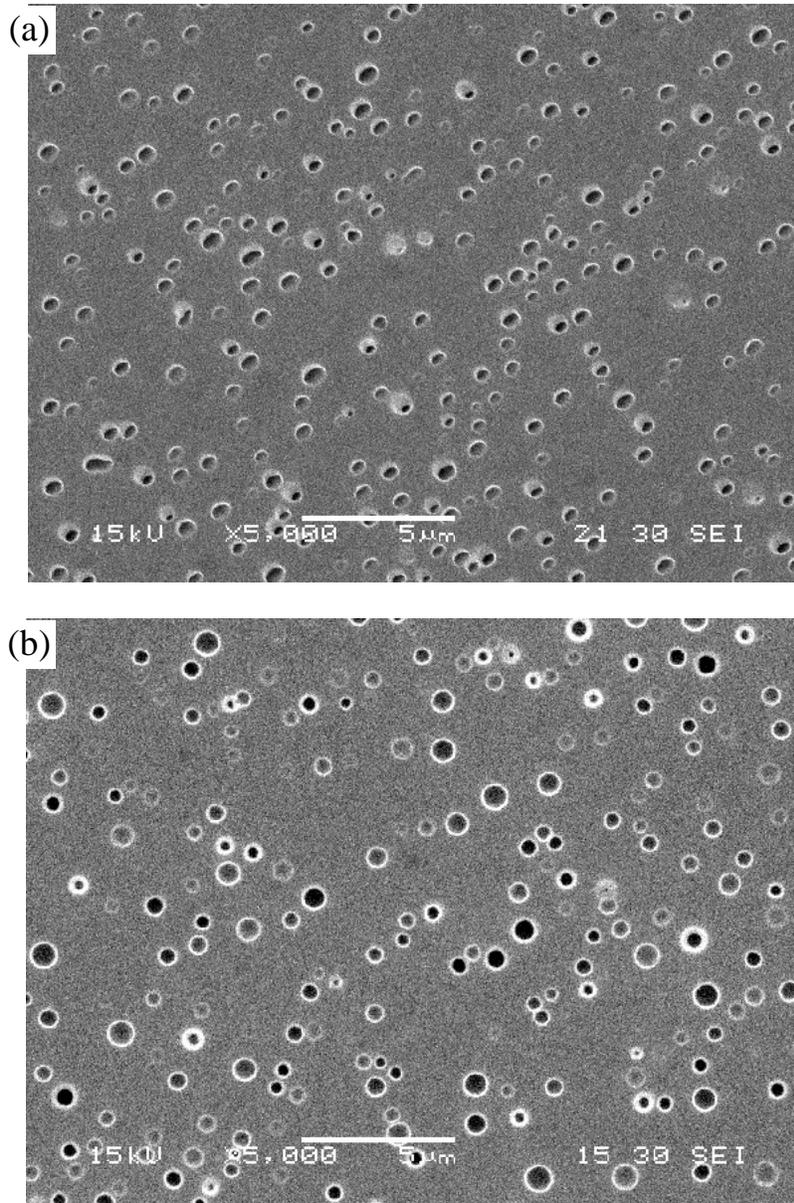

**Figure 7.** SEM micrographs for the blend with 30 wt % PIB5 obtained at different monomer conversions; (a) $p = 0.63$, (b) $p = 0.88$.

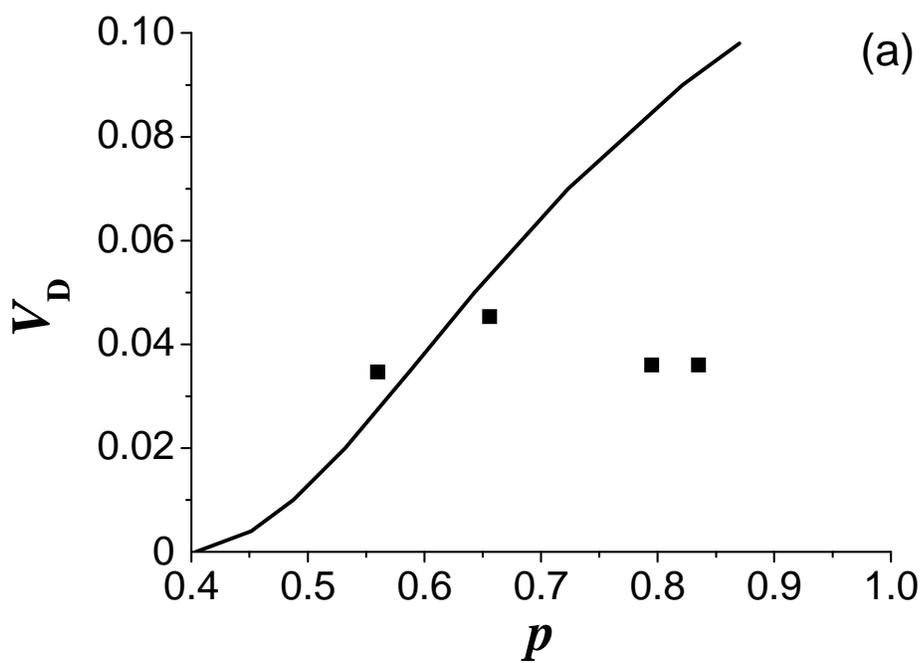

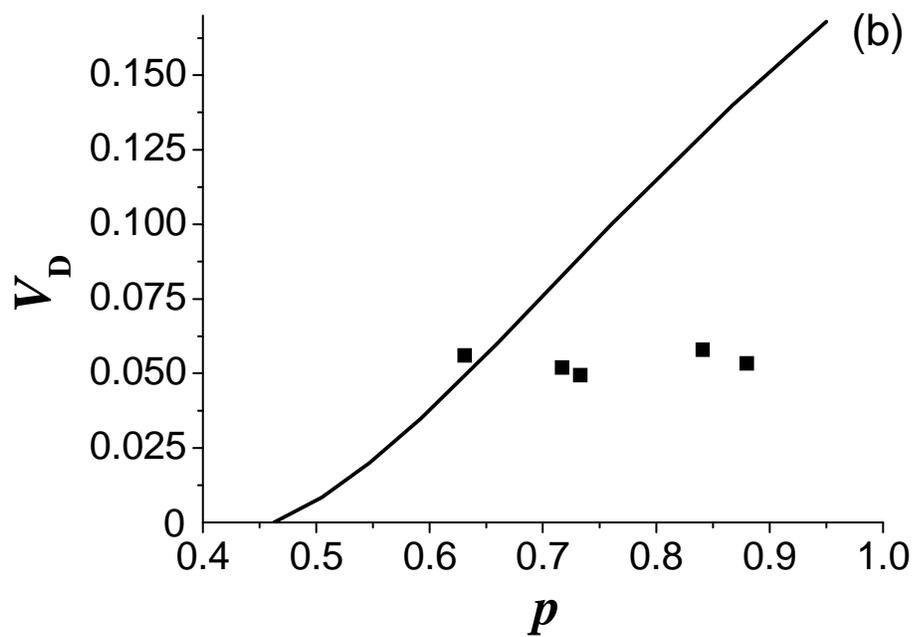

**Figure 8.** Volume fraction of dispersed phase predicted by the thermodynamic model assuming instantaneous equilibrium (curve) compared with experimental values obtained by SEM; (a) 15 wt % PIB30, (b) 30 wt% PIB5.

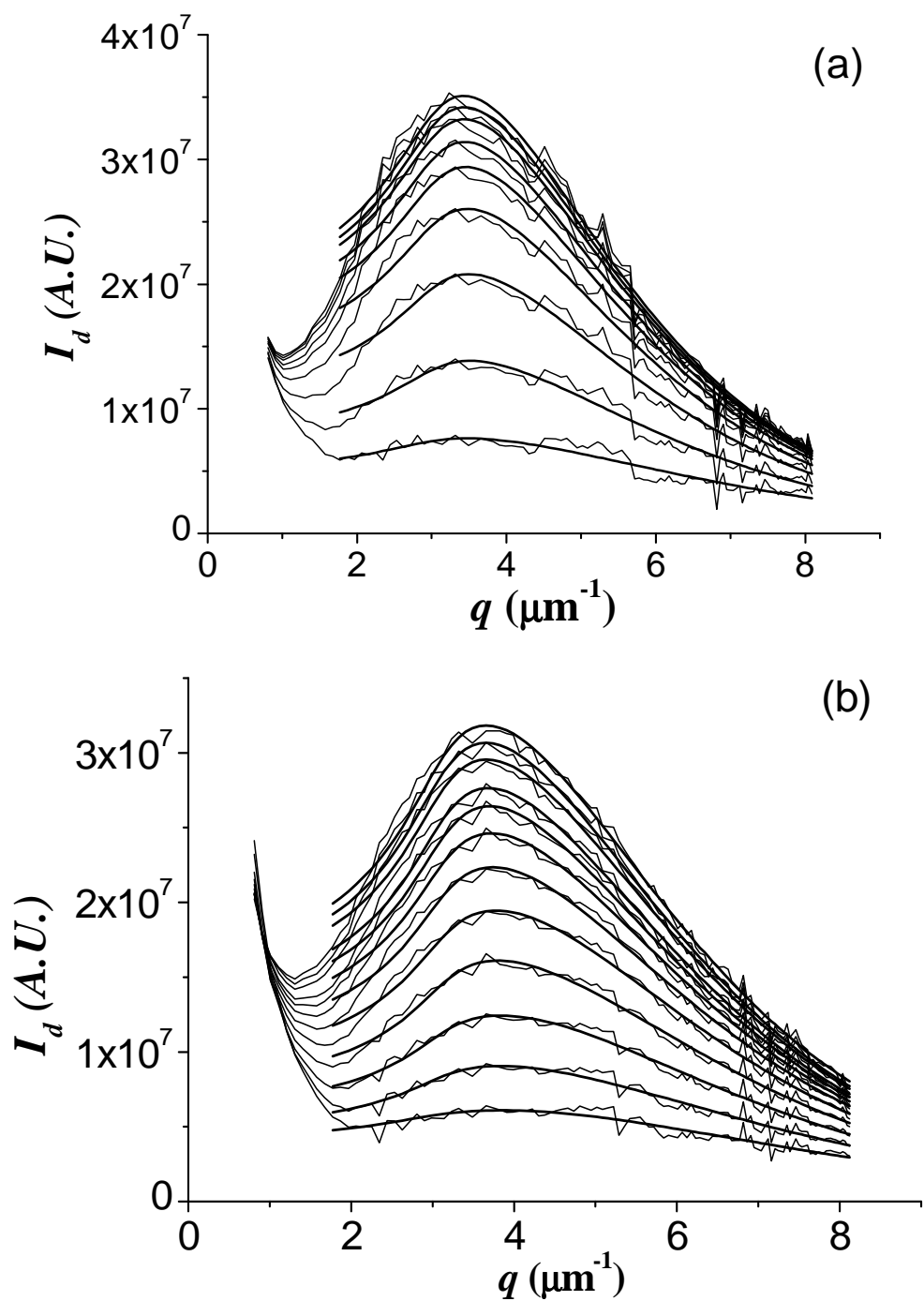

**Figure 9.** Light scattering spectra plotted as intensity (arbitrary units) as a function of the modulus of the scattering vector $q$. Non-smooth curves are experimental results and smooth curves represent the fitting of the LS model; (a) 15 wt % PIB30 for the following monomer conversions: 0.488, 0.541, 0.578, 0.61, 0.64, 0.667, 0.713, 0.753, 0.827; (b) 30 wt % PIB5 for the following monomer conversions: 0.512, 0.536, 0.558, 0.58, 0.601, 0.621, 0.641, 0.66, 0.678, 0.712, 0.788, 0.881.

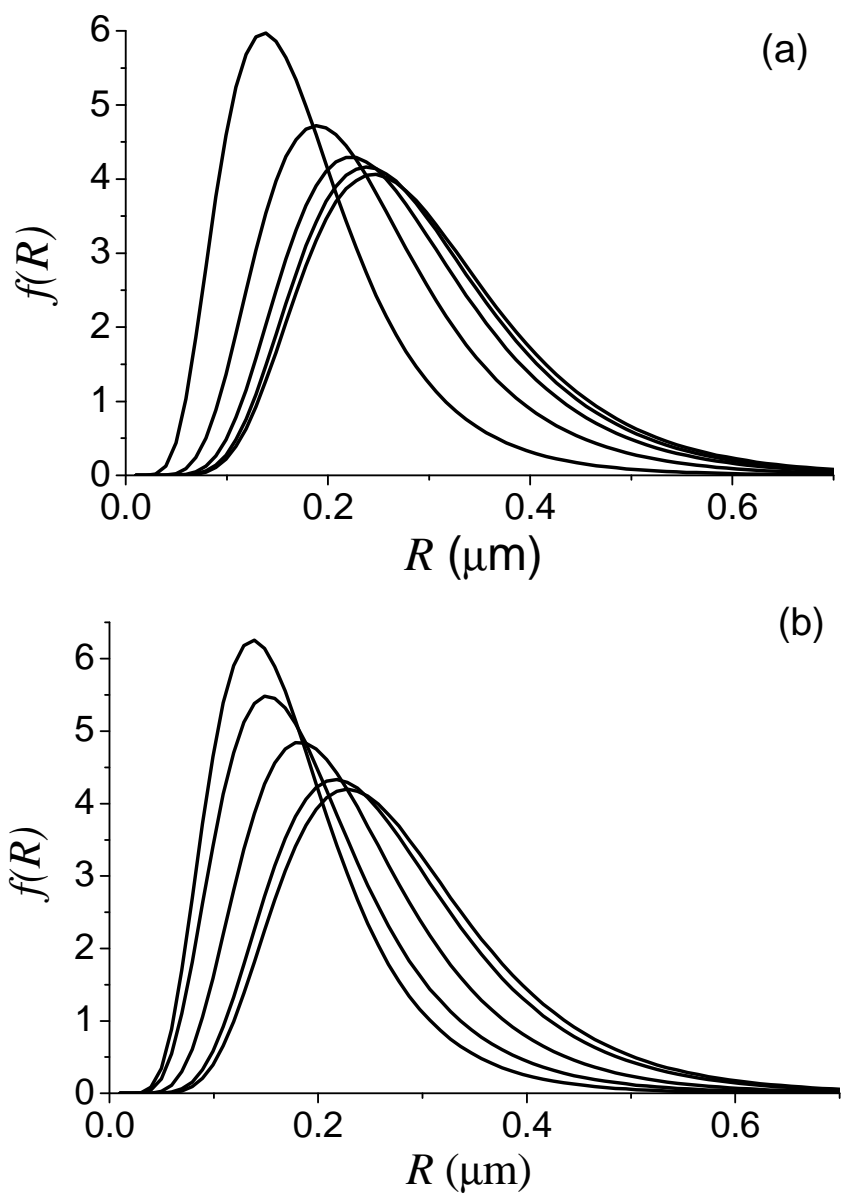

**Figure 10.** Evolution of particle size distributions calculated by the LS model for the two blends; (a) 15 wt % PIB30 for the following monomer conversions: 0.488, 0.541, 0.578, 0.61, 0.813; (b) 30 wt % PIB5 for the following monomer conversions: 0.512, 0.536, 0.558, 0.621 and 0.80.

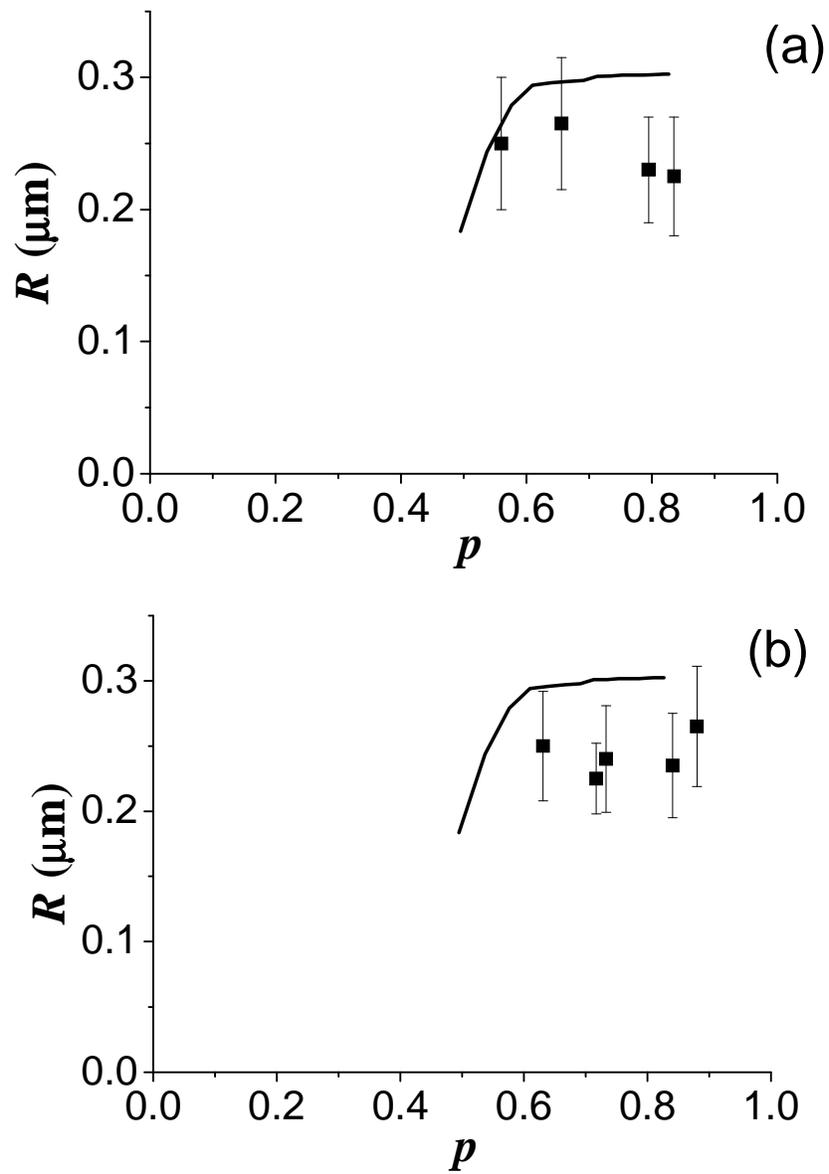

**Figure 11.** Comparison of average sizes calculated from the log-normal distribution with experimental values obtained from SEM micrographs; (a) 15 wt % PIB30, (b) 30 wt % PIB5.